\documentstyle[12pt]{article}

\begin{document}

%%%%%%%1111111111222222222233333333334444444444555555555566666666667777777777

\newcommand{\be}[1]{\begin{equation}\label{#1}}
\newcommand{\beq}{\begin{equation}}
\newcommand{\ee}{\end{equation}}
\newcommand{\beqn}[1]{\begin{eqnarray}\label{#1}}
\newcommand{\eeqn}{\end{eqnarray}}
\newcommand{\bd}{\begin{displaymath}}
\newcommand{\ed}{\end{displaymath}}
\newcommand{\mat}[4]{\left(\begin{array}{cc}{#1}&{#2}\\{#3}&{#4}\end{array}
\right)}
\newcommand{\matr}[9]{\left(\begin{array}{ccc}{#1}&{#2}&{#3}\\
{#4}&{#5}&{#6}\\{#7}&{#8}&{#9}\end{array}\right)}
\def\simlt{\mathrel{\lower2.5pt\vbox{\lineskip=0pt\baselineskip=0pt
           \hbox{$<$}\hbox{$\sim$}}}}
\def\simgt{\mathrel{\lower2.5pt\vbox{\lineskip=0pt\baselineskip=0pt
           \hbox{$>$}\hbox{$\sim$}}}}
\def\unity{{\hbox{1\kern-.8mm l}}}
\def\epr{E^\prime}
\def\al{\alpha}
\def\ga{\gamma}
\def\Ga{\Gamma}
\def\om{\omega}
\def\OM{\Omega}
\def\la{\lambda}
\def\La{\Lambda}
\newcommand{\eps}{\varepsilon}
\def\ep{\epsilon}
\newcommand{\ov}{\overline}
\renewcommand{\to}{\rightarrow}
\def\mcirc{{\stackrel{o}{m}}}
\newcommand{\bx}{\bar{\rm X}} 
\newcommand{\wx}{{\rm X}} 
\newcommand{\bv}{\bar{\rm V}} 
\newcommand{\wv}{{\rm V}} 
\newcommand{\tl}{\tilde{l}} 
\newcommand{\tq}{\tilde{q}} 
\newcommand{\tuc}{\tilde{u}_c} 
\newcommand{\tdc}{\tilde{d}_c} 
\newcommand{\tec}{\tilde{e}_c} 
\newcommand{\TQ}{\tilde{Q}} 
\newcommand{\TU}{\tilde{U}}
\newcommand{\TE}{\tilde{E}} 
\newcommand{\TUC}{\tilde{U}_c} 
\newcommand{\TEC}{\tilde{E}_c} 
\newcommand{\TQC}{\tilde{Q}_c} 
%\renewcommand{\thefootnote}{\fnsymbol{footnote}}
%%%%%%%%%%%%%%%%%%%%%%%%%%%%%%%%%%%%%%%%%%%%%%%%%%%%%%%

%\begin{titlepage} 

\begin{flushright} 
hep-ph/9605400 \\ 
INFN-FE-02/96 \\ 
May 1996 
\end{flushright}

\vspace{2cm}

\begin{center}

{\Large \bf Weak Mixing Angles as Dynamical Degrees of Freedom}
\vspace{1cm} 

{\large 
Alexei Anselm$^a$ ~ and ~ Zurab Berezhiani$^{b,c}$ } \\ [3mm]

$^a$ {\it Petersburg Nuclear Physics Institute, Gatchina, \\ 
188350 St.Petersburg, Russia} \\
\vspace{1mm}
$^b$ {\it INFN Sezione di Ferrara, 44100 Ferrara, Italy } \\
$^c$ {\it Institute of Physics, Georgian Academy of Sciences, \\
380077 Tbilisi, Georgia}

\end{center} 

\vspace{2cm}

\centerline{\bf abstract} 
\vspace{2mm} 

%\noindent 
In an analogy to the case of axion, which 
converts the $\Theta$-angle into a dynamical degree of freedom,  
we are trying to imagine a situation where the quark mixing angles 
turn out to be dynamical degrees of freedom (pseudo-Goldstone bosons), 
and their vacuum expectation values are obtained from the minimization 
of the vacuum energy. 
We present an explicit supersymmetric model with horizontal symmetry, 
where such a mechanism can be realized. 
It implies one relation between the quark masses and the CKM 
mixing angles: $s_{13}s_{23}/s_{12}=(m_s/m_b)^2$, which is fulfilled 
within present experimental accuracy. We believe, however, that 
the idea might be more general than this concrete model, 
and it can be implemented in more profound frameworks. 

% \maketitle

%\end{titlepage} 

\newpage

\section{Introduction}

The problem of CP violation in strong interaction, so-called 
$\Theta$-problem, can be most naturally resolved by the introduction
of the axion field which converts $\Theta$ parameter into a dynamical
degree of freedom \cite{PQ,axion,Kim}.

The pseudo-Goldstone boson, axion, is related to the chiral, flavour
nonchanging, transformations of quarks: global $U(1)_{\rm PQ}$ 
symmetry by Peccei and Quinn \cite{PQ}. This symmetry group can be 
extended to the rotations including a change between different 
generations.  Such a generalization of the Peccei-Quinn symmetry would
lead to the appearance of a set of Goldstone bosons -- familons
\cite{fam}.

Axion, being massless at the classical level, 
acquires small mass due to quantum corrections -- more precisely, 
due to axial anomaly -- and thus becomes a pseudo-Goldstone boson. 
The axion field acquires the vacuum expectation value (v.e.v.) 
which fixes the effective value of the $\Theta$ parameter. 
Namely, the minimum of the vacuum energy corresponds to
$\Theta=0$  resulting in the absence of strong CP violation.

The v.e.v.'s of the familons are unfixed as  long as the familons
remain true Goldstone bosons. However, though they cannot acquire
masses by the same mechanism as the axion, 
%related to the axial anomaly, 
they can nevertheless have small masses due to the explicit
breaking of the corresponding symmetry, possibly through the
radiative corrections.
If this happens the v.e.v.'s of the familon fields would fix the
mixing angles of the quarks, i.e. the Cabibbo--\-Kobayashi--\-Maskawa
(CKM) matrix, in the same way as the v.e.v. of the axion field fixes
the $\Theta$ parameter.

In other words, we are trying to imagine a situation where the quark 
mixing angles turn out to be dynamical degrees of freedom
(pseudo-\-Gold\-stone bosons) and  their vacuum expectation values
are obtained from the minimum of the vacuum energy.
We shall present an explicit example how such a mechanism can be 
realized. We believe, however, that the idea might be
more general than the concrete model described in this paper.

The complex, generally non-\-hermitian mass matrices of the up and 
down quarks can always be represented in the form:
\begin{equation}\label{a1}
M_{u}=U^\dagger_R M^{u}_{\rm diag} U_L\ , \qquad 
M_{d}=V^\dagger_R M^{d}_{\rm diag}V_L\ ,
\end{equation}
where 
\begin{equation}\label{a2} 
M^{u}_{\rm diag}= {\rm diag}(m_u,m_c,m_t)\ , \qquad 
M^{d}_{\rm diag}= {\rm diag}(m_d,m_s,m_b)\ ,
\end{equation} 
and $U_R,U_L,V_R,V_L$ are the unitary matrices 
%In the standard model the latter serve to diagonalize the (generally 
%non-\-hermitian) matrices of the Yukawa coupling constants.
%
%The matrices $U_L,V_L$ and $U_R,V_R$ 
which connect the quark mass eigenstates with the symmetry states 
("current quarks") as the latter appear in the Lagrangian. 
Evidently, the products $M_{u}^\dagger M_{u}$ and $M_{d}^\dagger M_{d}$ 
transform under the left-\-hand\-ed rotations of quarks:
\begin{equation}\label{a3}
M_{u}^\dagger M_{u} = U^\dagger_L (M^{u}_{\rm diag})^2 U_L\ , ~~~~~~~~
M_{d}^\dagger M_{d} = V^\dagger_L (M^{d}_{\rm diag})^2 V_L\ ,
\end{equation}
while $M_{u}M_{u}^\dagger$ and $M_{d}M_{d}^\dagger$ are related to the
right-hand\-ed rotations: 
\begin{equation}\label{a4}
M_{u} M_{u}^\dagger = U^\dagger_R (M^{u}_{\rm diag})^2 U_R\ , ~~~~~~~~
M_{d} M_{d}^\dagger = V^\dagger_R (M^{d}_{\rm diag})^2 V_R\ .
\end{equation}

The mixing of left-handed quarks in weak interaction is given by the 
CKM matrix $K_L = U_LV_L^\dagger$.   
The analogous matrix for the right-handed quarks, 
$K_R = U_R V_R^\dagger$, has not much physical sense in the 
absence of the right-handed weak interactions. 

Assume now that the fermion masses are actually the v.e.v.'s 
%vacuum expectation values 
of certain fields. It can be a very natural situation
that the minimum of the relevant Higgs potential, at least in the tree
approximation, would fix only the eigenvalues of $M_{u}$ and $M_{d}$
(i.e. the quark masses in $M^{u}_{\rm diag}$ and $M^{d}_{\rm diag}$) 
while the matrices $U_R,U_L,V_R,V_L$ would remain undefined.
To have this property it is sufficient that the Higgs potential would 
contain only the terms depending on the quantities 
${\rm Tr}(M_{u}^\dagger M_{u})$ and ${\rm Tr}(M_{d}^\dagger M_{d})$, 
but not on $M_{u}$ or $M_{d}$ themselves. (Of course,
we actually have in mind the appropriate Higgs fields whose v.e.v.'s
represent $M_{u}$ and $M_{d}$).

Furthermore, the potential in tree approximation may depend, or not
depend, on the structures 
\begin{equation}\label{a5}
%{\cal V}^L_{\rm eff} \sim 
{\rm Tr} \left[M_{u}^\dagger M_{u} 
M^\dagger_{d} M_{d} \right] = {\rm Tr} \left[ 
(M^{u}_{\rm diag})^2 K_L (M^{d}_{\rm diag})^2 K_L^\dagger \right]\ ,
\end{equation}
or the structures with a reverse order of $M_{u,d}$ and
$M_{u,d}^\dagger$: 
\begin{equation}\label{a6}
%{\cal V}^R_{\rm eff} \sim 
{\rm Tr} \left[ M_{u} M_{u}^\dagger 
M_{d} M_{d}^\dagger \right] = {\rm Tr} \left[
(M^{u}_{\rm diag})^2 K_R (M^{d}_{\rm diag})^2 K_R^\dagger \right]\ . 
\end{equation}
%
%If it does, the explicit dependence on the CKM matrix
%\begin{equation}
%$K\ =\ U_LV_L^\dagger $ 
%\end{equation}
%appears. 
If it does not, the dependence on the CKM matrix $K_L$ will anyway 
appear when the usual weak interaction is taken into 
account.\footnote{
In the absence of weak interactions, the total chiral symmetry of 
massless quarks would be 
$U(3)_{u_R}\times  U(3)_{u_L}\times U(3)_{d_R}\times U(3)_{d_L}$, 
so the scalars inducing the quark mass matrices $M_u$ and $M_d$ 
are respectively in representations $(3,\bar3,1,1)$ and  
$(1,1,3,\bar3)$. Clearly, no mixed structures like (\ref{a5}) 
or (\ref{a6}) are allowed by this $U(3)^4$ global symmetry. 
However, weak interactions are not invariant against independent 
rotations of $u_L$ and $d_L$ states and thus 
%reduce $U(3)_{u_L}\times U(3)_{d_L}$ factor to common $U(3)_{q_L}$, 
the term (\ref{a5}) becomes possible. } 
Indeed, when the quark masses are fixed, radiative corrections 
from weak interaction will lead to the contribution to the 
effective potential exactly of the form (\ref{a5}) through the loop 
diagram shown in Fig. 1. Indeed, it is straightforward to 
see that this diagram yields the contribution to the effective 
potential which depends on $K_L$:
\begin{equation}\label{a7}
{\cal V}_{\rm eff}\sim
\sum^3_{i,j=1}|K_{Lij}|^2 m^2_{u_i}m^2_{d_j}\ .
\end{equation}
where $m_{u_i}=(M^{u}_{\rm diag})_{ii}$, 
$m_{d_j}=(M^{d}_{\rm diag})_{jj}$ are the masses of the up and
down quarks. Due to Eq. (\ref{a3}), this expression exactly 
coincides with (\ref{a5}).

One can ask whether it is reasonable to assume the absence of the
contribution (\ref{a5}) to the effective potential in the tree
approximation if it anyway appears through the 
%convergent 
diagrams of Fig. 1? Clearly, the tree potential should include 
counterterm of the same structure.
The situation is analogous to a one of the pioneering work by Coleman
and Weinberg \cite{CW}. For a fixed, not too large value of a
cutoff the contribution of the loop diagrams is smaller than the
value of the tree potential. The smallness of the loop contribution
should be attributed then to the additional powers of the
dimensionless coupling constant.

In the standard model $SU(2)\times U(1)$ 
the left-handed quarks $q_{Li}= (u_L,d_L)_i$ 
transform as the doublets of the electroweak symmetry while the 
right-handed quarks $u_{Ri}$, $d_{Ri}$ are the weak singlets 
($i=1,2,3$ is a family index). 
%In the following we will use the left-handed basis of 
%fermions, and instead of right-handed states we consider their 
%antiparticles $u^c_{Li}=C\bar{u}_{Ri}$ and $d^c_{Li}=C\bar{d}_{Ri}$, 
%were $C$ is a charge conjugation matrix. 
The quark masses emerge via the Yukawa couplings
%(we skip index $L$ since we use only left-\-hand\-ed fermions):  
\begin{equation}\label{a8}  
{\cal L}_{\rm Yuk} = 
G_u^{ij}~ \bar{u}_{Ri} q_{Lj} \tilde{H} ~ + ~ 
G_d^{ij}~ \bar{d}_{Ri} q_{Lj} H 
\end{equation} 
where $H=(H^+,H^0)$ is the standard Higgs doublet with the v.e.v. 
$\langle H^0 \rangle = v$ ($v=(2\sqrt{2} G_F)^{-1/2}= 174$ GeV), 
and the Yukawa coupling constants $G_{u}^{ij}$ and $G_{d}^{ij}$ 
are $3\times 3$ complex matrices. The quark mass matrices are: 
\begin{equation}\label{a9} 
M_u=G_u v , ~~~~~~~~ M_d=G_d v 
\end{equation} 
Therefore, actually these are Yukawa coupling constants which we 
treat as dynamical degrees of freedom, assuming that they are 
given by v.e.v.'s of certain fields. In particular, 
we assume that the eigenvalues of the matrices $G_{u}$ and $G_{d}$, 
i.e. the values of quark masses, are frozen by the requirement
of the minimum of the tree-level potential of these fields. 
In what follows, they will be treated as fixed constants. 
At the same time the CKM matrix is related to a
set of dynamical degrees of freedom, the angles which enter the CKM
matrix are the v.e.v.'s of the pseudo-\-Gold\-stone bosons similar to
axion, and their values should be determined by the minimum of 
effective potential at the radiative level. In other words, we 
minimize the energy of the ground state with respect to the form 
of the unitary matrices in (\ref{a1}). 

In Figs. 2,3 we show the loop diagrams induced due to the Yukawa 
couplings (\ref{a8}). 
3-loop diagrams of Fig. 2 in fact contribute to the vacuum energy, 
and they all have a  
structure $\sim\Lambda^4 {\rm Tr}[G_u^\dagger G_u G_d^\dagger G_d]$, 
where $\Lambda$ is a cutoff scale 
(for the moment we omit the loop factors).   
The quadratically divergent 2-loop diagrams like the one of Fig. 3, 
where we deliberately inserted v.e.v.'s of Higgs $H$, give a structure 
$\sim v^2 \Lambda^2{\rm Tr} [G_u^\dagger G_u G_d^\dagger G_d]$. 
For the fixed Yukawa matrices this diagram represents a contribution 
to the Higgs doublet mass 
(among the other quadratically divergent contributions). However, 
for the given value of $H$ it can be treated as an effective potential 
term for the Yukawa degrees of freedom. 

Clearly, the quadratic divergency in the diagram of Fig. 3 will be 
removed as soon as one considers the supersymmetric theory \cite{SUSY}. 
In the case of unbroken supersymmetry the radiative corrections 
of Figs. 2,3 are exactly vanishing. 
Once supersymmetry is broken at the scale $m_S$ which can be from 
few hundred GeV to few TeV (i.e. roughly $m_S\sim v$), the contribution  
of Fig. 3 becomes $\sim m_S^2 v^2$. 
On the other hand, in the supersymmetric 
theory the vacuum energy is in general quadratically divergent. 
Hence,  contribution of diagrams in Fig. 2 
should be $\sim m_S^2\Lambda^2$.  
For the special choice of structures of the supersymmetry 
breaking soft terms the quadratic divergency can be removed 
also in the vacuum energy \cite{EDS}, in which case the contribution 
of Fig. 2 would become $\sim m_S^4$. However, in the rest of the 
paper we will not consider this specific case. 

Thus, in the context of our discussion the diagrams of Fig. 1 and 
Figs. 2,3, in spite of different degree of their divergency, are very 
similar: in fact, they all reproduce the structure (\ref{a5}).  
The insertion of the Higgs v.e.v. 
%$H^0_1,\widetilde H^0_2$ 
in the diagram of Fig. 3 ensures that the quarks become massive, 
which fact was implicitly assumed in the diagram of Fig. 1. 
One can say that the diagram of Fig. 1 has
been calculated after the spontaneous symmetry breaking had already
taken place while the diagram of Fig. 3 is used before the symmetry
breaking occurs. The $W$ exchange in Fig. 1 is
altered to the exchange of the charged Higgs boson in Fig. 3, 
which stays now instead of the longitudinal $W$ boson. 
Indeed, if the gauge coupling constant goes to zero, then the contribution
of the diagram of Fig. 1 does not vanish, as it may seem at first
glance, if one would substitute $M_W\sim gv$. This concur with the
non-\-va\-nish\-ing  contribution of the diagram of Fig. 3.

To summarize, we assume that the eigenvalues of the mass matrices
$M_{u}$ and $M_{d}$, i.e. quark masses (\ref{a2}), are frozen by the 
requirement of the minimum of the tree-level potential. 
In what follows, they will be treated as fixed constants. 
At the same time the CKM matrix is related to a
set of dynamical degrees of freedom, the angles which enter the CKM
matrix are the v.e.v.'s of the pseudo-\-Gold\-stone bosons similar to
axion. 
We have argued that if the term fixing the relative orientation of
$M_u^\dagger M_u$ and $M_d^\dagger M_d$ is absent in a tree-level 
potential, then it is induced radiatively. 
%by the electroweak interactions. 
The effective potential, which must fix the v.e.v.'s of these fields, 
i.e. weak mixing angles, should at least contain the term  
(\ref{a5}) since this structure is dictated by
usual weak interactions.

On the other hand, if the structure (\ref{a6}) is absent in the 
tree-level potential, then it will not emerge after the 
radiative corrections as long as the
right-\-hand\-ed fermions do not have $SU(2)_R$ gauge interactions
and the related Yukawa couplings. That means that the relative
rotation angles of the right-\-hand\-ed quarks correspond to the true
Goldstone degrees of freedom. They will not be considered in this
paper as well as any other true Goldstones -- familons.

According to our scenario the v.e.v.'s of the pseudo-Goldstone bosons
(so to say, "pseudo-\-fa\-mi\-lons") fix the mixing angles, 
just like the v.e.v. of axion field fixes the 
$\Theta$-angle.\footnote{
%Before passing to more detailed considerations, 
Let us remark that besides the axion, 
there exists one more example when a rotation angle is actually 
a dynamical degree of freedom. This is an usual pion field 
corresponding to chiral rotations of quarks in the isotopic
space. In the chiral limit when the current masses of the light
quarks vanish, $m_u=m_d=0$, pions are true Goldstone bosons and their
v.e.v.'s are undefined. For $m_u,m_d\neq0$ pions become
pseudo-\-Gold\-stones and their v.e.v.'s turn out to be zero,
$\langle \pi^a\rangle=0$ (in a "reference frame" in O(4) chiral space
determined by the condition $\langle\bar qq\rangle\neq0,$
$\langle\bar q\tau^a\gamma_5q\rangle=0$).
} 
The masses of these bosons are related to the absolute value of the
loop contribution of the type of (\ref{a5}) to the effective potential. 
It is very difficult to estimate the value of ${\cal V}_{\rm eff}$. 
However, we shall argue below that if the cutoff is actually given 
by the TeV scale supersymmetry, 
%with a soft breaking scale $m_S\sim v$, 
the pseudo-\-fa\-mi\-lon masses can be in the MeV range. 

Unfortunately, the potential which contains only a term (\ref{a5}) does
not lead to nontrivial mixing angles: the angles all vanish if it 
enters ${\cal V}_{\rm eff}$ with a negative coefficient, or 
all equal to $\pi/2$ if this coefficient is positive. 
Therefore, we are obliged to add some different structures. 
This problem will be discussed in the next sections.

The paper is organized as follows. In Section 2 we continue to
discuss a possible symmetry structure of the effective potential 
which could provide a non-trivial solution for the CKM angles. 
The effective potential is presented in a special parametrization. 
In Section 3 a concrete model based on a chiral horizontal $SU(3)_H$  
symmetry of generations is considered, in which naturally emerges 
the general structure of the effective potential assumed in Section 2. 
In Section 4 we discuss a toy model with two generations of fermions. 
This example allows to explain the underlying physical mechanism,  
and also serves as a technical tool in considering the more complicated 
realistic case of three generations. The latter case is considered 
in Section 5.  The mixing angles are found and one physical relation 
between the angles is established (see Eq. (\ref{kiko})), which 
is satisfied within the present experimental accuracy. In Section 6 
we estimate a range of possible masses of pseudo-\-Gold\-stone
bosons related to the CKM mixing angles. Some concluding remarks are
given at the end of the paper.

\section{Effective potential for the CKM matrix}

In the standard model the gauge interactions of fermions obey the 
global symmetry related to the inter-family transformations of 
the different types of fermions: 
$U(3)_{u_R}\times U(3)_{d_R}\times U(3)_{q_L}$. 
As it was explained above, one can imply by the fermion "masses" 
the appropriate Higgs v.e.v.'s, so  
the Higgses responsible for $M_u$ and $M_d$  
transform respectively as $(3,1,\bar3)$ and $(1,3,\bar3)$  
representations of this $U(3)^3$ group. 

The simplest assumption is that the 
effective potential in the tree approximation depends only on the traces 
of the powers of $M_{u}^\dagger M_{u}$ and $M_{d}^\dagger M_{d}$.  
%but not $m^{(u)}$ and $m^{(d)}$ themselves. 
This would leave all the rotation matrices in Eq. (\ref{a1}) to be 
the Goldstone degrees of freedom. Then radiative corrections induce 
the structure (\ref{a5}) in the effective potential which lifts the 
vacuum degeneracy with respect to CKM angles, and thus fixes their 
values. However, as we have already mentioned, if only the 
structure (\ref{a5}) is present in the effective potential, 
the mixing angles are trivial. 

%Of course, the latter also violates SU(3) symmetries
%when non-\-equ\-al masses of the quarks are fixed (see (ref{a8})).

Next in simplicity would be an assumption that the third generation
of the fermions is somewhat different from the first two.
Qualitatively we can express this by considering the terms 
${\rm Tr}[ M_{u}^\dagger M_{u} \lambda_8]$ and 
${\rm Tr}[ M_{d}^\dagger M_{d} \lambda_8]$.
We assume that these terms indeed appear together with the term 
(\ref{a5}) in the effective potential, which now acquires a form: 
%which fixes the rotation angles.
\begin{equation}\label{b1}
{\cal V}_{\rm eff} = 
A\,{\rm Tr} \left[M_{u}^\dagger M_{u} \lambda_8 \right] + 
B\,{\rm Tr} \left[M_{d}^\dagger M_{d} \lambda_8 \right] + 
C\,{\rm Tr} \left[M_{u}^\dagger M_{u} M_{d}^\dagger M_{d} \right]\ ,
\end{equation}
where $A,B$ and $C$ are some unknown constants.

In the next section we shall present an explicit model which has
exactly these properties. In this model all the structures 
in (\ref{b1}) emerge at the radiative level due to the
spontaneous symmetry breaking rather than in explicit manner. 
%violation of the SU(3) symmetries. 

%The structures (\ref{a10}) together with (\ref{a2}) result in the following
%potential:
%\begin{eqnarray}
%{\cal V}_{\rm eff} &= & 
%A\,{\rm Tr} \left[M^{u+} M^{u} \lambda_8 \right] + 
%B\,{\rm Tr} \left[M^{d+} M^{d} \lambda_8 \right] \nonumber\\
%&+& C\,{\rm Tr} \left[M^{u+} M^{u} M^{d+} M^{d} \right]\ ,
%\end{eqnarray}

Substituting  Eqs. (\ref{a3}) in (\ref{b1}), one obtains:
\begin{eqnarray}\label{b2}
{\cal V}_{\rm eff} = 
A\,{\rm Tr}\left[(M^{u}_{\rm diag})^2 U \lambda_8 U^\dagger  \right] + 
B\,{\rm Tr}\left[(M^{d}_{\rm diag})^2 V \lambda_8 V^\dagger)\right] 
\nonumber \\
+ C\,{\rm Tr} \left[(M^{u}_{\rm diag})^2 K  
(M^{d}_{\rm diag})^2 K^\dagger \right] 
%~~~~~~~~ K_L=U_LV_L^\dagger \ .
\end{eqnarray} 
where  $U\equiv U_L$, $V\equiv V_L$, and $K=UV^\dagger$. 
This basic expression can be reorganized in the following way. 
First we notice that the matrix $\lambda_8\sim {\rm diag}(1,1-2)$ 
can be changed to $\lambda_0\sim {\rm diag}(0,0,1)$, since the terms 
(\ref{b2}) with the unit matrix $I$ instead of $\lambda_8$ does not 
depend on $U$ and $V$. 
Then, without lose of generality, one can also substract 
from $(M^{u}_{\rm diag})^2$ 
%$(M^{u}_{\rm diag})^2=(m^2_t,m^2_c,m^2_u)\to(m^2_t-m^2_u,m^2_c-m^2_u,0)$ 
and $(M^{d}_{\rm diag})^2$ 
%$(M_{d}^{\rm diag})^2 =(m^2_b,m^2_s,m^2_d)\to(m^2_b-m^2_d,m^2_s-m^2_d,0)$, 
respectively the unit matrices $m_u^2I$ and $m_d^2I$.
% there is again no dependence on $U_L,V_L$ and $K_L$. 
Therefore, the expression (\ref{b2}) can be presented in the form:
\begin{equation}\label{b3}
{\cal V}_{\rm eff} =  
A\sum_{i=2,3}|U_{i3}|^2 \bar{m}^2_{u_i} + 
B\sum_{i=2,3}|V_{i3}|^2 \bar{m}^2_{d_i} +
C\sum_{i,j=2,3}|K_{ij}|^2 \bar{m}^2_{u_i} \bar{m}^2_{d_i} \ , 
%~~~~ K=UV^+\ ,
\end{equation}
%\begin{eqnarray}
%{\cal V}_{\rm eff} &=& 
%A\sum_{i=2,3}|U_{i3}|^2 \bar{m}^2_{u_i} + 
%B\sum_{i=2,3}|V_{i3}|^2 \bar{m}^2_{d_i} \nonumber \\
%&+& C\sum_{ij=2,3}|K_{ij}|^2 \bar{m}^2_{u_i} \bar{m}^2_{d_i} \ , 
%~~~~ K=UV^+\ ,
%\end{eqnarray}
where  
%$U\equiv U_L,\;V\equiv V_L$, $K=UV^\dagger$ and
\begin{eqnarray}\label{b4}
&& 
\bar{m}^2_{c} =m^2_c-m^2_u \simeq m^2_c\ , \qquad 
\bar{m}^2_{t} =m^2_t-m^2_u \simeq m^2_t\ , \nonumber \\ 
&&
\bar{m}^2_{s} =m^2_s-m^2_d \simeq m^2_s\ , \qquad 
\bar{m}^2_{b} =m^2_b-m^2_d \simeq m^2_b\ .
\end{eqnarray}
Of course, $A,B,C$ in Eq. (\ref{b3}) are not the same as in 
Eq. (\ref{b2}). In the following for $\bar{m}^2_{c,s,t,b}$ we use 
their approximate values (\ref{b4}): 

We shall parametrize the $3\times3$ unitary matrices $U$ and $V$ by
three consecutive unitary transformations acting between the  
$(1,2)$, $(2,3)$ and $(1,2)$ generations:
\begin{eqnarray}\label{b5}
U&=&U_{12}U_{23}U'_{12}\ , ~~~~~~~ V=V_{12}V_{23}V'_{12}\nonumber \\
&&\\
K&=&UV^\dagger=U_{12}U_{23}S_{12}V^\dagger_{23}V^\dagger_{12}\ , ~~~~
S_{12}=U'_{12}V'^\dagger_{12}\ . \nonumber
\end{eqnarray}
The advantages of this parametrization are obvious. First, since the
matrices $U'_{12}$ and $V'_{12}$ commute with $\lambda_8$, they drop
out in two first terms for the potential in the expression (\ref{b2}).
Second, only their product, $S_{12}$, remains in the third term. To
introduce the necessary 6 independent phases in $U$ and $V$ we
include three phases in each matrix $U_{12}$, $U_{23}$: 
%$V_{12}$,$V_{23}$:
\begin{eqnarray}\label{b6}
&&
U_{12} = \left(\begin{array}{ccc} e^{i\alpha_{12}}\cos\Theta_{12} &
e^{i\beta_{12}}\sin\Theta_{12} & 0 \\
-e^{i\gamma_{12}}\sin\Theta_{12} & e^{i\delta_{12}}\cos\Theta_{12} &
0\\ 0 & 0 & 1 \end{array}\right) ,  ~~~~ 
\alpha_{12}-\beta_{12} = \gamma_{12}-\delta_{12} ,   \nonumber \\
&& \\ 
&& 
U_{23} = \left(\begin{array}{ccc} 1  & 0 & 0 \\
0 & e^{i\alpha_{23}}\cos\Theta_{23} & e^{i\beta_{23}}
\sin\Theta_{23} \\ 0 & -e^{i\gamma_{23}}\sin\Theta_{23} &
e^{i\gamma_{23}}\cos\Theta_{23}\end{array}\right) , ~~~~
\alpha_{23}-\beta_{23}=\gamma_{23}-\delta_{23}\ , \nonumber
\end{eqnarray}
and analogously for $V_{12}$ and $V_{23}$, but with the change of
notations: $\Theta_{12}\to\widetilde\Theta_{12}$,
$\alpha_{12}\to \widetilde\alpha_{12},\ldots$,  $\Theta_{23}\to
\widetilde\Theta_{23}$, $\alpha_{23}\to\widetilde\alpha_{23},\ldots$,
etc.

The matrices $U'_{12}$ and $V'_{12}$ actually can be chosen
orthogonal. Only their product, $S_{12}=U'_{12}V'^\dagger_{12}$, 
enters the expression (\ref{b2}) for ${\cal V}_{\rm eff}$. 
We parametrize:
\begin{equation}\label{b7}
S_{12}=\left(\begin{array}{ccc} \cos\omega & \sin\omega & 0 \\ 
-\sin\omega & \cos\omega & 0 \\ 
0  & 0  &1  \end{array}\right)
\end{equation}

The straightforward calculation of the effective potential (\ref{b3})
shows that it depends only on three combinations of the phases, namely:
\begin{eqnarray}\label{b8}
\Phi_1&=&\alpha_{23}-\beta_{23}-\widetilde\alpha_{23}+\widetilde
\beta_{23}, \nonumber \\
\Phi_2&=&\alpha_{23}-\beta_{23}-\widetilde\alpha_{12}+
\widetilde\beta_{12}+ \widetilde\beta_{23}, \\
\Phi_3&=&\alpha_{12}-\beta_{12}- \beta_{23}
-\widetilde\alpha_{23}+\widetilde\beta_{23}\ . \nonumber
\end{eqnarray}
Indeed, substituting (\ref{b5}) in Eq. (\ref{b3}) we obtain: 
\begin{eqnarray}\label{b9}
&&
{\cal V}_{\rm eff} = Am^2_t\cos^2\Theta_{23} 
+ Am^2_c\cos^2\Theta_{12}\sin^2 \Theta_{23}   
+ Bm^2_b\cos^2\widetilde\Theta_{23}    \nonumber \\ 
&& 
+ Bm^2_s\cos^2\widetilde\Theta_{12}\sin^2\widetilde\Theta_{23} 
+ Cm^2_t m^2_b \bigg|\cos\Theta_{23}\cos\widetilde\Theta_{23}
+\sin\Theta_{23}\sin\widetilde\Theta_{23} 
\cos\omega e^{i\Phi_1}\bigg|^2    \nonumber \\
&& 
+ Cm^2_t m^2_s
\bigg|\cos\widetilde\Theta_{12}\left(\cos\Theta_{23}
\sin\widetilde\Theta_{23}-\sin\Theta_{23}\cos\widetilde\Theta_{23}
\cos\omega e^{i\Phi_1}\right)    \nonumber \\
&& 
- \sin\Theta_{23}\sin\widetilde\Theta_{12}
\sin\omega e^{i\Phi_2}\bigg|^2    
+ Cm^2_c m^2_b\bigg|\cos\Theta_{12}\big(\sin\Theta_{23}
\cos\widetilde\Theta_{23}     \nonumber \\
&& 
- \cos\Theta_{23}\sin\widetilde\Theta_{23}\cos\omega e^{i\Phi_1}\big)   
+ \sin\widetilde\Theta_{23}\sin\Theta_{12}
\sin\omega e^{i\Phi_3}\bigg|^2 \nonumber \\ 
&&
+ Cm^2_c m^2_s
\bigg|\cos\Theta_{12}\cos\widetilde\Theta_{12}
\left(\sin\Theta_{23}\sin\widetilde\Theta_{23}+\cos\Theta_{23}
\cos\widetilde\Theta_{23}\cos\omega e^{i\Phi_1}\right) \nonumber \\ 
&& 
+ \cos\Theta_{12}\cos\Theta_{23}\sin\widetilde\Theta_{12}
\sin\omega e^{i\Phi_2} - \sin\Theta_{12}\cos\widetilde\Theta_{12}
\cos\widetilde\Theta_{23}\sin\omega e^{i\Phi_3} \nonumber \\ 
&&
+ \sin\Theta_{12}\sin\widetilde\Theta_{12}\cos\omega 
e^{-i\Phi_1+i\Phi_2+i\Phi_3}\bigg|^2\ .
\end{eqnarray}

In the following, for parametrization of the CKM matrix we adopt 
the "standard" choice advocated by the Particle Data Group \cite{PDG}: 
\begin{equation}\label{CKM} 
K = \matr{c_{12}c_{13}}{s_{12}c_{13}}{s_{13}e^{-i\delta}}
{-s_{12}c_{23}-c_{12}s_{23}s_{13}e^{i\delta}}
{c_{12}c_{23}-s_{12}s_{23}s_{13}e^{i\delta}}{s_{23}c_{13}} 
{s_{12}s_{23}-c_{12}c_{23}s_{13}e^{i\delta}}
{-c_{12}s_{23}-s_{12}c_{23}s_{13}e^{i\delta}}
{c_{23}c_{13}} 
\end{equation}
where $s_{12}=\sin\vartheta_{12}$, $c_{ij}=\cos\vartheta_{12}$, etc., 
and $\delta$ is the CP violating phase. 

\section{The model}

To carry out explicitly the program which was outlined in the
previous sections we use the model with the chiral horizontal 
$SU(3)_H$ symmetry between the generations \cite{su3h}. 
In this model left-\-hand\-ed quarks $q^\alpha_{Li}=(u,d)_{Li}$ 
transform as triplets of $SU(3)_H$, whereas the right-\-hand\-ed
ones $u_R^i$ and $d_R^i$ are anti-\-tri\-plets ($i=1,2,3$ is an 
index of generations).  In this paper we will not consider leptons, 
though clearly they can be included in a strightforward way.

Using only left-handed fields and, consequently, $u^c_L=C\bar u_R$
and $d^c_L=C\bar d_R$ instead of $u_R$ and $d_R$, we can write the
simplest Yukawa couplings which can lead to the appearance of the
quark masses in the following form 
(we skip subscript $L$ since we use only the left-handed fermions): 
\begin{equation}\label{c1} 
% L=\epsilon_{\alpha\beta}(u^c_iCq^\alpha_j)\chi^{ij,\beta}
%+(d^c_iCq^\alpha_j)\xi^{\alpha*}_{ij}+H.C.
u^c_i q^\alpha_j \epsilon_{\alpha\beta} H_2^{ij,\beta} ~ + ~ 
d^c_i q^\alpha_j \epsilon_{\alpha\beta} H_1^{ij,\beta} 
\end{equation}
where $H_2^{ij,\beta}$ and $H_1^{ij,\beta}$ represent a set of 
the Higgs doublets of $SU(2)\times U(1)$ (index $\alpha,\beta=1,2$) 
which simoultaneously transform as $\bar 6$ or $3$ under $SU(3)_H$. 
%\begin{eqnarray}
%H_2^{ij}\sim(3,2) & \mbox{ or } & (\bar6,2)\ ,\nonumber\\
%H_1^{ij}\sim(3,2) & \mbox{ or } & (\bar6,2)\ .
%\end{eqnarray}
The problem with the couplings (\ref{c1}) is that they lead to the 
flavour changing neutral currents (FCNC), as always happens when more 
than one Higgs doublet gives masses to the quarks with the same charge 
of different generations \cite{FCNC}. It is not easy to suppress 
naturally these currents \cite{su3h}.

One way to overcome this difficulty is to change the fields 
$H_2^{ij,\alpha}$ and $H_1^{ij,\alpha}$ in Eq. (\ref{c1}) by the 
products of the Higgs fields which are transformed trivially by each 
of the groups $SU(3)_H$ and $SU(2)\times U(1)$ \cite{FN}.
Namely, let us put \cite{ZB85}: 
\begin{equation}\label{c2} 
H_2^{ij,\beta}=\frac{\chi^{ij}\cdot H^\beta_2}M\ , \qquad
H_1^{ij,\beta}=\frac{\xi^{ij}\cdot H^\beta_1}M \ ,
\end{equation}
where $\chi^{ij}$ and $\xi^{ij}$  are transformed as $3$ or $\bar6$ 
of $SU(3)_H$ and are singlets of $SU(2)\times U(1)$, while $H_{1,2}$ 
are doublets of $SU(2)\times U(1)$ and the $SU(3)_H$ singlets. 
$M$ is the mass parameter which is introduced to preserve the right 
dimension of the fields. In other words, we consider 
%Substituting (3.3) into (3.2) we get the
non-\-re\-nor\-ma\-liz\-able interactions 
\begin{equation}\label{c3}
\frac{\chi^{ij}}M 
\cdot u^c_i q^\alpha_j \epsilon_{\alpha\beta} H^\beta_2
~ + ~ \frac{\xi^{ij}}M 
\cdot d^c_i q^\alpha_j \epsilon_{\alpha\beta} H^\beta_1 
\end{equation}
One sees that for large enough $M$ the interaction of the quarks with
the scalars can be made as weak as necessary whereas the usual values
of the masses of the fermions can be set up by the appropriate choice
of the v.e.v.'s of $\chi^{ij}$ and $\xi^{ij}$. In fact the ratios
$\langle\chi^{ij}\rangle/M$ and $\langle\xi^{ij}\rangle/M$ are
nothing but the matrices $G_u$ and $G_d$ of the Yukawa coupling constants 
in the standard model.\footnote{Notice, that 
actual global chiral symmetry of the terms (\ref{c3}) is 
$U(3)_H=SU(3)_H\times U(1)_H$, where $U(1)_H$ is related to 
phase transformation $q,u^c,d^c \to e^{i\varphi}q,u^c,d^c$, 
$\chi,\xi \to e^{-2i\varphi}\chi,\xi$. Thus the fermion mass 
hierarchy in fact is a reflection of the v.e.v.'s hierarchy in 
the chiral $U(3)_H$ symmetry breaking 
$U(3)_H \to U(2)_H\to U(1)_H \to {\em nothing}$. 
In fact, $U(1)_H$ can serve as the Peccei-Quinn symmetry unless 
it is explicitly broken in the potential of $\chi$ and $\xi$ 
\cite{ZB85}.  }

%$H_1=H$ and $H_2=\tilde{H}={\rm i}\tau_2 H^\ast$, where 
%In the framework of the minimal supersymmetric standard model (MSSM) 
%eq. (\ref{a1}) denotes the superpotential terms with $H_1$ and $H_2$ 
%being two different Higgs superfields, respectively  
% with v.e.v.'s $v_1=v\cos\beta$ and $v_2=v\sin\beta$.

As a matter of fact, what we actually have in mind in considering 
this model, is a supersymmetric theory. In other words, 
$q$, $u^c$ and $d^c$ are chiral superfields of quarks, 
$H_{1,2}$ are the MSSM Higgs doublets and $\chi$ and $\xi$ 
are `horizontal' Higgs superfields breaking the $SU(3)_H$ 
symmetry.\footnote{In the following, as it is usually adopted, 
we distinguish the fermion and Higgs superfields by their 
matter parity, negative for fermions and positive for Higgses.} 
For the completeness of the theory, in principle 
one has to introduce also the Higgs superfields $\bar{\chi}$ and 
$\bar{\xi}$ in representations conjugated to $\chi$ and $\xi$, 
but these do not play a relevant role in our further considerations.  
Eq. (\ref{c3}) actually are the superpotential terms responsible 
for quark masses. 

There are different ways to justify the appearance of the
non-renormalizable interactions (\ref{c3}). Maybe the most natural and
simplest way is to introduce the additional vector-like set of 
heavy fermions \cite{FN}, namely, the weak isosinglets transforming 
as triplet representation of $SU(3)_H$ 
%for both the left-\-hand\-ed and right-\-hand\-ed components 
\cite{ZB85}. 
In other words, per each generation we introduce the left 
chiral ($SU(2)$-singlet) partners $U_i, U^i_c$ and $D_i, D^i_c$ 
($i=1,2,3$),  with the same electric and colour charges as 
$u,u^c,d,d^c$ but with the following transformation properties under 
$SU(3)_H$:
\begin{equation}\label{c4}
U_i,D_i\sim 3 \ , \qquad U^i_c,D^i_c\sim \bar3 \ .
\end{equation}

The assignment (\ref{c4}) allows the large mass terms 
("survival hypothesis") for the states $U,U^c$ and $D,D^c$: 
\begin{equation}\label{c5}
M(U_c^i U_i)\ , \qquad M(D_c^i D_i)\ ,
\end{equation}
as well as their couplings
\begin{equation}\label{c6}
(U_c^{i} U_j)\Sigma^j_i \ ,\qquad (D_c^{i} D_j)\Sigma^j_i \ ,
\end{equation}
with the scalar $\Sigma$ in an adjoint (octet) representation 
of $SU(3)_H$: $\Sigma\sim 8$. 
It is natural to assume that due to a tree-level 
potential, $\Sigma$ develops the v.e.v. proportional to $\lambda_8$: 
$\langle\Sigma\rangle \sim {\rm diag} (1,1,-2)$.  
Of course, the mass parameter $M$ can be
different in $U$ and $D$ mass terms as well as the coupling constants
for the two structures of (\ref{c6}). However, this is irrelevant 
for our discussion.

The Yukawa couplings which lead now to the masses of the light
quarks are:
\begin{equation}\label{c7}
(u^c_i U_j)\chi^{ij}\ , \qquad (d^c_i D_j)\xi^{ij}
\end{equation}
and
\begin{equation}\label{c8}
(U_c^{i} q^\alpha_i) \epsilon_{\alpha\beta} H^\beta_2\ , \qquad
(D_c^{i} q^\alpha_i) \epsilon_{\alpha\beta} H^\beta_1 \ ,
\end{equation} 
where we absorbe the coupling constants into the Higgs fields.

All Yukawa couplings of the light and heavy fermions, 
in the basis of $(u,d,U,D)$ and $(u^c,d^c,U^c,D^c)$ states,  
can be presented in the form of a field-dependent mass matrix:
\begin{equation}\label{c9}
{\cal M} = 
{\left(\begin{array}{cccc} 0&0&H^0_2 & H^-_1\\ 
0 & 0 & H^+_2 & H^0_1 \\ \chi&0&M+\Sigma &0\\ 0&\xi&0&
M+\Sigma \end{array}\right)} . 
\end{equation}
The constant mass matrix emerges
when the scalars are changed by their v.e.v.'s.

The effective non-renormalizable Lagrangian (\ref{c3}) emerges through 
the diagrams of the type of Fig. 4 
%(iterated in $M$ and summed up) 
in the limit $M\gg \chi,\xi, \Sigma$. Hence, mass matrices of the 
up- and down-quarks are connected to the v.e.v.'s of $\chi$ and $\xi$:
\begin{equation}\label{c10}
M_{u}= G_u \langle H^0_2\rangle ,~~~ 
G_u^{ij}=\frac{\langle\chi^{ij}\rangle}M\ ; \qquad
M_{d}= G_d \langle H^0_1\rangle , ~~~ 
G_d^{ij}=\frac{\langle\xi^{ij}\rangle}M\ .
\end{equation}
Each of these v.e.v.'s can be brought to diagonal form: 
$\langle\chi\rangle \sim {\rm diag}(\chi_1,\chi_2,\chi_3)$  and 
$\langle\xi\rangle \sim {\rm diag}(\xi_1,\xi_2,\xi_3)$, 
so that in the `seesaw' limit $\chi_i,\xi_i\ll M$ 
the quark masses in (\ref{a2}) are essentially the 
ratios $\chi_i/M$ and $\xi_i/M$.   
Clearly, the large value of the top mass requires 
$\chi_{3} \sim M$, whereas other v.e.v.'s should 
be much smaller than $M$. Actually, for the top mass one has 
to use more precise formula (see e.g. in ref. \cite{top}) 
%in diagonalization of the total mass matrix (\ref{c9}) 
rather than the one given in seesaw limit, Eq. (\ref{c10}). 
However, this is not of principal 
importance for our consideration. In addition, since our model has 
a rather illustrative character, we do not take 
into account the renormalization running of masses from the scale of 
the horizontal symmetry down to the electroweak scale. 

In spirit of our proposal, we assume that a tree-level superpotential 
of  $\chi,\xi$ and $\Sigma$ contains only the self-interaction 
terms of these fields like ${\rm Tr}\,(\bar{\chi}\chi)$, 
${\rm Tr}\,(\bar{\chi}\chi\bar{\chi}\chi)$, ${\rm Tr}\,(\Sigma^2)$,
${\rm Tr}\,(\Sigma^3)$, etc., but does not contain crossing terms 
like ${\rm Tr}\,(\bar{\chi}\xi)$, ${\rm Tr}\,(\bar{\chi}\chi\Sigma)$, etc. 
At this level, potential can fix a shape of v.e.v.'s of each of these 
fields, but the relative orientation of the v.e.v.'s 
of $\chi$, $\xi$ and $\Sigma$ remains unfix. 
In other words, superpotential has a global symmetry 
$SU(3)_\chi\times SU(3)_\xi\times SU(3)_\Sigma$ 
related to independent unitary transformations 
of $\chi,\xi$ and $\Sigma$. 

The Yukawa terms do not respect the $SU(3)^3$ global symmetry, 
and hence radiative corrections should violate it also in the 
Higgs potential. Nevertheless, 
if supersymmetry is unbroken, no additional structures 
will emerge in radiative corrections and thus the CKM angles 
would remain the true Goldstone modes. However, once supersymmetry 
is broken, radiative corrections will become effective. They 
remove the vacuum degeneracy and give rise to certain terms 
in the effective potential which link these scalars to each other. 
The soft supersymmetry breaking can be accounted by the 
spurion superfield $z=m_S\theta^2$ ($\bar{z}=m_S\bar{\theta}^2$), 
where $m_S\sim v$ is the soft mass scale. 
Then the desired structures (\ref{b1}) could emerge from D-terms
\begin{equation}\label{spurion}
\left[{\rm Tr}\,(\chi^+\chi\Sigma)z\bar{z}\right]_D\ , \qquad 
\left[{\rm Tr}\,(\xi^+\xi\Sigma)z\bar{z}\right]_D\ , \qquad 
\left[{\rm Tr}(\chi^+\chi\xi^+\xi)z\bar{z}\right]_D \ .
\end{equation}

The first two terms in (\ref{spurion}) indeed emerge from 
the one-loop supergraphs shown in Fig. 5, after inserting 
the spurion fields into the internal lines or vertices.  
By taking into account that $\langle \Sigma\rangle \sim \lambda_8$, 
these terms would immediately translate into the 
first two terms of the effective potential (\ref{b1}).
Clearly, from the similar diagrams (with insertion of $\Sigma$ 
instead of mass entry $M$), also the terms like 
${\rm Tr}\,(\xi^+\xi\Sigma^+\Sigma)$ will be induced. 
However, these in fact do not create new structures in (\ref{b1}), 
since $\lambda_8^2$ is a combination of the unit matrix and 
$\lambda_8$ itself. 

The third term in (\ref{spurion}) emerges from the 3-loop 
graph of Fig. 6, where under the non-renormalizable 
vertices we actually imply the effective operators induced 
by the heavy (with mass $M$) fermion exchanges as in Figs. 4. 
For the momenta smaller than $M$, when our theory effectively 
reduces to the non-renormalizable operators (\ref{c3}), 
this graphs effectively reduce to the ones given in Fig. 2, 
which (in supersymmetric case) are quadratically divergent. 
Therefore, $M$ actually acts as a cutoff scale and the contribution 
of this diagram is $\sim m_S^2 M^2\,{\rm Tr}(G_u^+G_uG_d^+G_d)$. 
%which gives the third term in (\ref{b1}).    

Thus, after the supersymmetry breaking the following terms 
emerge in the effective potential of the scalars $\chi,\xi$ and 
$\Sigma$ (the loop factors are omitted):  
\begin{equation}\label{c11}
\frac{m_S^2}{M}\,{\rm Tr}\,(\chi^+\chi\Sigma)\ , \qquad 
\frac{m_S^2}{M}\,{\rm Tr}\,(\xi^+\xi\Sigma)\ , \qquad 
\frac{m_S^2}{M^2}\,{\rm Tr}(\chi^+\chi\xi^+\xi) \ .
\end{equation}
which after substituting the basic tree-level v.e.v.'s 
of these fields reduce to the ${\cal V}_{\rm eff}$ of the structure 
given by Eq. (\ref{b1}). The same order of magnitude of all three 
terms can be achieved by properly chosen  values of  
$\langle\Sigma\rangle \ll M$ and of the coupling constants 
in Eq. (\ref{c6}). 

The following comment is in order. For the vacuum expectation
value of $\Sigma$ we have assumed that actually only one component 
of the octet does not vanish: $\langle\Sigma\rangle\sim\lambda_8$. 
Such a solution indeed emerge in an unique way from the 
superpotential of $\Sigma$, $W(\Sigma)=m\Sigma^2 + \Sigma^3$.  
One can expect the similar properties for sextets and triplets. 
Certainly, there is no reason for $\chi$ or
$\xi$ to have several non-\-va\-nish\-ing components but rather one
non-\-va\-nish\-ing  eigenvalue for each matrices $\xi^{ij}$ and
$\chi^{ij}$. In other words, their v.e.v.'s can be (independently) 
rotated to the form $\sim(0,0,1)$. 
As a result, only one up- and one down-\-qu\-ark would acquire 
the non-zero masses. However, the other non-zero 
eigenvalues in $\xi$ and $\chi$ can be induced by their interactions 
to the other set of superfields $\xi'$ and $\chi'$ in some 
representations of $SU(3)_H$  which themselves do not couple to 
fermions.\footnote{Alternatively, 
one could introduce several Higgs fields in the place of $\chi$ or
$\xi$, say a set of sextets and triplets for each, with v.e.v.'s 
on different components. What we actually mean then by $\chi^{ij}$ 
and $\xi^{ij}$ are in fact the relevant combinations of these fields 
in which they couple to fermions. } 
It this way all quarks can get masses. 
Furthermore our assumption is that the Higgs 
superpotential is organized in such a way that it is invariant under 
the separate $SU(3)$ rotations of all fields composing 
$\xi$, $\chi$, or $\Sigma$. In other words, we assume 
that it has a form  
\begin{equation}\label{c12}
W= W(\xi,\xi') + W(\chi,\chi') + W(\Sigma) 
\end{equation} 
respecting the accidental global symmetry 
$SU(3)_\chi\times SU(3)_\xi\times SU(3)_\Sigma$ 
related to the independent unitary transformations 
of $\chi(\chi')$, $\xi(\xi')$ and $\Sigma$. Then the relative $SU(3)$
orientation of $\xi$ and $\chi$ as well as their relative orientation
to $\Sigma$ will be fixed by the loop contributions to 
${\cal V}_{\rm eff}$ leading to the expression (\ref{b1}).

Concluding, in the case of the exact supersymmetry the structures 
(\ref{c11}), once they are absent in tree-level potential, 
would not appear in radiative corrections. Broken supersymmetry 
allows to generate such terms, however suppresses their values 
so that they are proportional to $m_S^2$.  
They appear in effective potential with values $\sim m_S^2 M^2$,  
much smaller than typical size ($\sim M^4$) of the tree-level 
terms like $(\chi^+\chi)^2$ etc. 
(certainly, there is also an additional suppression due to the 
loop factors).
 Therefore, pseudo-familons are indeed light, with masses 
$<m_S$.\footnote{Let us remark that there are other interesting 
examples when the flat directions of the supersymmetric 
theory give rise to light states in the particle spectrum. 
One popular example is, e.g. when the MSSM 
Higgs doublets appear as the pseudo-Goldstone bosons of the 
accidental global symmetry \cite{GIFT}. } 
Below we shall try to estimate the magnitude of the loop diagrams of
Figs. 5,6, and hence the values of the pseudo-familon masses. 
At the moment we confine ourselves by the observation that
the model indeed leads to the ${\cal V}_{\rm eff}$ of the structure 
given by Eq. (\ref{b1}).

\section{A toy model for two generations}

In this section we shall consider the non-realistic model of two
generations of fermions: $u_i=(c,t)$, $d_i=(s,b)$. The purpose of
this exercise is twofold: first, this simplified version very well
illustrates the physics related to our approach. Second, below we
shall use the results of this section in treating the realistic case
of three generations. In the expression (\ref{b2}) for the effective
potential we change $\lambda_8\to {\rm diag}(0,1)$.  

Let us parametrize the $2\times 2$ unitary matrices $U$ and $V$ as
\begin{eqnarray}\label{d1}
&&
U = \left(\begin{array}{cc} e^{i\alpha}\cos\Theta &
e^{i\beta}\sin\Theta \\ -e^{i\gamma}\sin\Theta &
e^{i\delta}\cos\Theta \end{array}\right)\ , 
~~~~~ \alpha -\beta = \gamma -\delta , \nonumber\\
&& ~~~~ \\   
&&
V = \left(\begin{array}{cc} 
e^{i\tilde\alpha}\cos\tilde\Theta & 
e^{i\tilde\beta}\sin\tilde\Theta \\ 
-e^{i\tilde\gamma}\sin\tilde\Theta &
e^{i\tilde\delta}\cos\tilde\Theta  \end{array}\right)\ , 
~~~~ \tilde\alpha-\tilde\beta=\tilde\gamma-\tilde\delta\ .  \nonumber
\end{eqnarray}
Then the same arguments
which were used to obtain Eq. (\ref{b4}) lead to the result:
%\begin{equation}\label{d1}
%{\cal V}_{\rm eff} = 
%A\bar{m}^2_t |U_{33}|^2 + B\bar{m}^2_b |V_{33}|^2 + 
%C \bar{m}^2_t \bar{m}^2_b \left|(UV^\dagger)_{33}\right|^2 \ .
%\end{equation}
%Then expression (\ref{d1}) takes the form:
\begin{eqnarray}\label{d2}
{\cal V}_{\rm eff} &=&
A m^2_t |U_{33}|^2 + Bm^2_b |V_{33}|^2 + 
Cm^2_t m^2_b \left|(UV^\dagger)_{33}\right|^2 = 
\nonumber \\ 
&=& A m^2_t \cos^2\Theta +B m^2_b \cos^2\tilde\Theta\ + 
Cm^2_t m^2_b \left|\cos\Theta\cos\tilde\Theta
+e^{i\Phi} \sin\Theta\sin\tilde\Theta \right|^2,  \nonumber \\
&& ~~~~~~~~
\Phi = \delta-\tilde\delta - \gamma +\tilde\gamma\ . 
\end{eqnarray}
The extremum of this potential corresponds to $\sin\Phi=0$, i.e.
$\Phi=0,\pi$. By allowing both signs for $\Theta,\tilde\Theta$, we can
choose $\Phi=0$ and simplify Eq. (\ref{d2}) in the following way, 
omitting an unessential additive constant: 
\begin{equation}\label{d3}
{\cal V}_{\rm eff}=\ a\cos\chi_1+b\cos\chi_2+c\cos\chi_3
\end{equation}
where $\chi_{1}=2\Theta$, $\chi_{2}=2\tilde\Theta$, 
$\chi_3 =2(\Theta-\tilde\Theta)$ and  
\begin{equation}\label{d4}
a=\frac12\,A m^2_t\ , b=\frac12\,B m^2_b\ , 
c=\frac12\,Cm^2_t m^2_b      \nonumber
\end{equation}

To gain some physical intuition it is useful to interprete the
expression (\ref{d3}) as a potential energy of the system of three 
interacting two-\-di\-men\-sion\-al unit vectors: $\vec n_0(1,0)$,  
$\vec n_1(\cos\chi_1,\sin\chi_1)$, $\vec n_2(\cos\chi_2,\sin\chi_2)$. 
In terms of these vectors
\begin{equation}\label{d5}
{\cal V}_{\rm eff}=\ a\,(\vec n_0\cdot\vec n_1) + 
b\,(\vec n_0\cdot\vec n_2) + c\, (\vec n_1\cdot\vec n_2)\ .
\end{equation}
Each positive coefficient (say $a>0$) describes the "repulsion" of
the corresponding pair of vectors. The minimum of the corresponding
term, i.e. $a(\vec n_0\cdot\vec n_1)$, is reached for 
$(\vec n_0\cdot\vec n_1)=-1$, i.e. $\chi_1=\pi$. Any negative 
coefficient (e.g. $b<0$) can be understood as an attraction, 
and the minimum corresponds
now to $(\vec n_0\cdot\vec n_2)=1$, i.e. $\chi_2=0$.

Clearly, if all three couplings are attractive: 
%It is clear that for all three attractions: 
$a,b,c<0$, then the absolute minimum of ${\cal V}_{\rm eff}$ 
is obtained when all three vectors $\vec n_{0,1,2}$ are parallel: 
$\chi_1=\chi_2=\chi_3=0$. 
If there are two repulsions and one attraction
(say, $a>0,b>0$ and $c<0$), then $\chi_1=\chi_2=\pi$, $\chi_3=0$: 
two vectors $\vec n_1$ and $\vec n_2$ are stuck to each other but 
oriented in the opposite direction to the third one $\vec n_0$. 

There are two cases when one can expect the nontrivial configuration
of the vectors and therefore nontrivial mixing angles (see Fig. 7)
First case corresponds to three repulsions, $a,b,c>0$. Second is
realized for two attractions and one repulsion 
(for example, $a,b<0$ and $c>0$). In this latter case vectors 
$\vec n_1$ and $\vec n_2$ are attracted to $\vec n_0$, but their 
mutual repulsion does not
allow them to stick to each other and to $\vec n_0$.

The equation $d{\cal V}_{\rm eff}/d\chi_{1,2}=0$, 
besides the trivial solution
$\sin\chi_1=\sin\chi_2=\sin\chi_3=0$, has a non-\-trivial one:
\begin{eqnarray}\label{d6}
\cos\chi_1 = \frac12\left(\frac{bc}{a^2}-\frac bc-\frac cb\right),\
\;\;\;
\cos\chi_2=\frac12\left(\frac{ac}{b^2}-\frac ac-\frac ca\right)\ ,
\nonumber \\
\cos\chi_3 = \frac12\left(\frac{ab}{c^2}-\frac ab-\frac ba\right).
\end{eqnarray}
To analyze these solutions, it is useful to write down the
expressions for $\sin^2\Theta_i$ 
(here we put $c=1$, i.e. rescale $a/c,b/c\to a,b)$: 
\begin{eqnarray}\label{d7}
\sin^2\Theta &=& \frac12(1-\cos\chi_1)=\frac1{4a^2b}
(ab+a+b)(ab+a-b)\ , \nonumber \\
\sin^2\tilde\Theta &=& \frac12(1-\cos\chi_2)=\frac1{4ab^2}
(ab+a+b)(ab-a+b)\ ,           \\
\sin^2(\Theta -\tilde\Theta)&=&\frac12(1-\cos\chi_3)=\frac1{4ab}
(ab+a+b)(a+b-ab)\ . \nonumber
\end{eqnarray}
The requirement that all $\sin^2\Theta_i$ should be positive leads to
the  region of the allowed values of $a$ and $b$ shown in Fig. 8
(for $c=1$ regions I and III, for $c=-1$ regions II and IV). 
It is also easy to prove that in this region $\sin^2\Theta_i<1$.

One can see from Fig. 8 that,   
in order to have nontrivial mixing angles for $c=1$, 
one should either have $a$ and $b$ both positive or both of them
negative. An additional feature is that for 
positive $a,b$, i.e. for the case of all three repulsions, for enough 
large values of $a$ and $b$, when  $ab-a-b>0$, there is no 
non-\-tri\-vi\-al mixing.
Consider, for example, the simple case when $a=b$. 
Then for $a>2$ the repulsion of $\vec n_1$ and $\vec n_2$ from 
$\vec n_0$ is so strong that they stick to each other
in the opposite direction to $\vec n_0$, in spite of repulsion 
between themselves. The value $a=b=2$ is sort of
a "threshold": for the smaller values of $a$ a small angle between
$\vec n_1$ and $\vec n_2$ appears. This angle grows as $a$ decreases
and reaches $120^\circ$ for $a=b=1$. In this latter symmetric case
(we remind that $c=1$) $\vec n_0$, $\vec n_1$ and $\vec n_2$ compose
a configuration with all the angles equal $120^\circ$.

The similar phenomenon has place in the case of $a,b<0$. If the
attraction of $\vec n_1$ and $\vec n_2$ to $\vec n_0$ is very strong, 
so that $ab+a+b>0$ $(e.g. a<-2$ for $a=b$), all three
vectors are stuck to each other. Only for smaller $|a|,|b|$ the
mixing angle appears. 

On the other hand, it is easy to show that when
$0\leq \sin^2\Theta_i \leq 1$, i.e. inside the regions of Fig. 8, 
the non-\-trivial solution (\ref{d6}) (or (\ref{d7})) leads to lower 
energy than the trivial ones. 
For example, for the case of
three repulsions, $a,b,c>0$, the magnitude of the effective potential 
(\ref{d3})  
%of ${\cal V}^{(ext)}_{\rm eff}$ 
for the $\cos\chi_i$ from (\ref{d6}) is the following: 
\begin{equation}\label{d8}
{\cal V}^{(ext)}_{\rm eff}=\ -a-b+c-\frac{(ac+bc-ab)^2}{2abc}\ , 
\end{equation}
which is always smaller as comparted to the magnitude 
${\cal V}^{(0)}_{\rm eff}=-a-b+c$ at the trivial extremum 
$\chi_{1,2}=\pi,\, \chi_3=0$.  

%corresponding to $\chi_1=\chi_2=\pi$ for
%$a,b,c>0$ or $\chi_1=\chi_2=0$ for $a,b<0$, $c>0$. 
%Thus  $V^{(ext)}_{eff}<V_{eff}(\chi_1=\pi,\, \chi_2=\pi,\, \chi_3=0)
%=-a-b+c$. 

For the case $a,b<0$ but $c>0$ we have:
\begin{equation}\label{d9}
{\cal V}^{(ext)}_{\rm eff}=\ a+b+c-\frac{(ab+bc+ac)^2}{2abc}\ .
\end{equation}
which is again less than the magnitude 
${\cal V}^{(0)}_{\rm eff}=a+b+c$ at the trivial extremum 
$\chi_{1,2,3}=0$.  
%Again $V^{(ext)}_{eff}<V_{eff}(\chi_1=\chi_2=\chi_3=0)=a+b+c$. 
In fact, the mixing angles outside the regions of Fig. 8 become 
trivial (zero or $\pi$) not because the energy of the trivial solution 
becomes lower than the energy corresponding formally to (\ref{d6}), 
but  solely because that there are no
physical solutions fulfilling the condition $0<\sin^2\Theta_i<1.$

\section{Three generations: fixing the CKM angles }

We now pass to the realistic case of three generations. The basic
expression for ${\cal V}_{\rm eff}$, which we use in what follows, is
given by Eq. (\ref{b9}).

The solution with $\sin\Phi_i=0$, $i=1,2,3$, is certainly an exact
extremum of ${\cal V}_{\rm eff}$. One can argue that it is
unlikely to have different extremums corresponding
to non-\-tri\-vi\-al phases. 

Let us first focus on the dependence of ${\cal V}_{\rm eff}$ on $\Phi_1$. 
The leading term is proportional to $m^2_t m^2_b$:
\begin{equation}\label{e1} 
{\cal V}_{\rm eff}\sim2Cm^2_tm^2_b\cos\Phi_1 \cos \Theta_{23}\cos
\widetilde\Theta_{23}\sin\Theta_{23}\sin\widetilde\Theta_{23}
\cos\omega \ .
\end{equation}
Though it is difficult to have a rigorous proof concerning the next
terms proportional to $m^2_tm^2_s$, $m^2_cm^2_b$ and $m^2_cm^2_s$
they seem to be negligible as compared to (\ref{d1}). For example, 
the term  $\sim m^2_tm^2_s\cdot\cos(\Phi_1-\Phi_2)$ contains even 
more sinuses then the leading contribution (\ref{e1}) and it is very 
difficult to imagine what could compensate the smallness of $m^2_s/m^2_b$.

Leaving only the contribution (\ref{e1}), we see that the non-trivial
solution of the equation $d{\cal V}_{\rm eff}/d\Phi_1=0$ 
(i.e. when $\Theta_{23}$, $\widetilde\Theta_{23}\neq0,\pi/2$, ~ 
$\Theta'_{12}\neq\pi/2$) implies that $\sin\Phi_1=0$.  
%\begin{equation}
%\frac{dV_{eff}}{d\Phi_1}=0\ \to\ \sin\Phi_1=0\ .
%\end{equation}
Thus, $\Phi_1=0$ or $\Phi_1=\pi$. We shall choose $\Phi_1=0$ and see
that this value corresponds to a minimum of ${\cal V}_{\rm eff}$. 
Indeeed, from (\ref{e1}) one has: 
\begin{equation}\label{e2}
\left(\frac{d^2{\cal V}}{d\Phi^2_1}\right)_{\Phi_1=0}=-2Cm^2_tm^2_b
\cos\Theta_{23}\cos\widetilde\Theta_{23}\sin\Theta_{23}\sin
\widetilde\Theta_{23}\cos\omega\ .
\end{equation}
For the solution described below: $C<0$, $\Theta_{23}$ and
$\widetilde\Theta_{23}$ are small and have the same sign, and 
$\omega=0$. Therefore $(d^2{\cal V}/d\Phi^2_1)_{\Phi_1=0}>0$.

Similar arguments are applicable to $\Phi_2$ and $\Phi_3$, after we
insert $\Phi_1=0$. We assume therefore that $\Phi_2=\Phi_3=0$.
For the solution given below $(\omega=0)$ we obtain analogously
to (\ref{e1}):  
\begin{eqnarray}\label{e3} 
{\cal V}_{\rm eff}(\Phi_2,\Phi_3)&\sim &
2Cm^2_cm^2_s\cos(\Phi_2+\Phi_3)\cos
\Theta_{12}\cos\widetilde\Theta_{12} \nonumber \\
& \times &\cos(\Theta_{23}-
\widetilde\Theta_{23})\sin\Theta_{12}\sin\widetilde\Theta_{12}.
\end{eqnarray}
Again $\Theta_{12}$ and $\widetilde\Theta_{12}$ are small and have
the same sign. Therefore, ${\cal V}_{\rm eff}$ has a minimum 
%in  $\Phi_2$ and $\Phi_3$ 
at $\Phi_2=\Phi_3=0$.

Next step is to find the minimum in $\omega$. Contrary to the
case of the phases $\Phi_i$, $\omega=0$ is not an exact
solution of the equation $d{\cal V}_{\rm eff}/d\omega=0$. However, 
if one neglects non-\-lead\-ing contributions and leaves only the 
term $\sim m^2_tm^2_b$, then $\omega=0$ is indeed an extremum. We
adopt this approximation and write down the simplified expression for
${\cal V}_{\rm eff}$ for $\Phi_1=\Phi_2=\Phi_3=\omega=0$:
\begin{eqnarray}\label{e4} 
{\cal V}_{\rm eff} = Am^2_t\cos^2\Theta_{23} 
+ Am^2_c\cos^2\Theta_{12} \sin^2\Theta_{23}  
+ Bm^2_b\cos^2 \widetilde\Theta_{23} \nonumber \\ 
+ Bm^2_s\cos^2 \widetilde\Theta_{12}\sin^2\widetilde\Theta_{23}   
+ Cm^2_tm^2_b \cos^2(\Theta_{23}-\widetilde\Theta_{23}) \nonumber \\ 
+ Cm^2_tm^2_s \cos^2\widetilde\Theta_{12}\sin^2(\Theta_{23} -
\widetilde\Theta_{23})     
%\nonumber \\
+ Cm^2_cm^2_b\cos^2\Theta_{12}\sin^2(\Theta_{23}-
\widetilde\Theta_{23})  \nonumber \\
+ Cm^2_cm^2_s\left[\sin\Theta_{12}\sin
\widetilde\Theta_{12}+\cos\Theta_{12}\cos\widetilde\Theta_{12}
\cos(\Theta_{23}-\widetilde\Theta_{23})\right]^2  
\end{eqnarray}

%\section{Three generations. Mixing angles}

To find explicitly the mixing angles we shall use the mass hierarchy
and the smallness of the mixing angles. With the accuracy of order of
$10^{-4}-10^{-5}$ the leading terms in Eq. (\ref{e4}) are: 
\begin{equation}\label{e5}
{\cal V}_{\rm eff}=Am^2_t\cos^2\Theta_{23}+Bm^2_b\cos^2
\widetilde\Theta_{23}+Cm^2_tm^2_b\cos^2(\Theta_{23}-
\widetilde\Theta_{23}) 
%= \nonumber \\ 
%2\left[ Am^2_t\cos2\Theta_{23} + Bm^2_b\cos2\widetilde\Theta_{23}
%+ Cm^2_tm^2_b\cos2(\Theta_{23}-\widetilde\Theta_{23})\right]
%+\mbox{const}
\end{equation}
Thus the problem of finding $\Theta_{23}$ and $\widetilde\Theta_{23}$
reduces to the two-\-ge\-ne\-ra\-tion case considered in Section 4.
As it was explained in this section, the only way to get the
non-\-tri\-vi\-al  mixing angles is to have two negative
and one positive coefficients in the expression for ${\cal V}_{\rm eff}$. 
If one rewrites (\ref{e5}) in the form of Eq. ({\ref{d5}) 
%\begin{equation}
%V_{eff}=2Am^2_t\cos2\Theta_{23}+2Bm^2_b\cos2\widetilde\Theta_{23}
%+2Cm^2_tm^2_b\cos2(\Theta_{23}-\widetilde\Theta_{23})+\mbox{ const}
%\end{equation}
and identifies the cosines of the double angles as the scalar products 
of the unit vectors: $\cos2\Theta_3=\vec n_1\cdot\vec n_0$, $\cos2
\widetilde\Theta_{23}=\vec n_2\cdot\vec n_0$, $\cos2(\Theta_{23} -
\widetilde\Theta_{23})=\vec n_1\cdot\vec n_2$, this case would
correspond to an attraction of a two pairs of the vectors and one
repulsion. We shall choose the situation shown in Fig. 7B corresponding
to $A<0$, $C<0$, $B>0$ when $\vec n_0$ and $\vec n_2$ are both attracted
to $\vec n_1$ but repulse from each other. Clearly, in this case the 
"gluey" vector $\vec n_1$ should be placed  between 
$\vec n_0$ and $\vec n_2$.

To use the results of Section 4 we rewrite the expression (\ref{e5}),
omitting an unessential additive constant, in the form
\begin{eqnarray}\label{e6}
\frac{{\cal V}_{\rm eff}}{2|C|m^2_tm^2_b}&=&-a\cos2\Theta_{23}+b\cos2
\widetilde\Theta_{23}-\cos2(\Theta_{23}-\widetilde\Theta_{23}) \\
a&=&\left|\frac A{Cm^2_b}\right|\ ,\ \ b=\left|\frac B{Cm^2_t}
\right|\ . \nonumber
\end{eqnarray}
In Eq. (\ref{d7}) $a$ and $b$ mean actually $a/c$ and $b/c$ with
$c=1$. Since the expression (\ref{e6}) differs from (\ref{d3}) 
by the change $a\to -a$ and $c\to-c=-1$, 
we can directly use the  expression (\ref{d7}) 
changing the ratios $a/c\to a/c$, $b/c\to-b/c$, i.e. 
$a\to a$, $b\to-b$. Thus we get:
\begin{eqnarray}\label{e7}
\sin^2\Theta_{23}&=&\frac1{4a^2b}(ab-a+b)(a+b-ab)\ ,\nonumber \\
\sin^2\widetilde\Theta_{23} &=&\frac1{4ab^2}(ab-a+b)(a+b+ab)\ , \\
\sin^2(\Theta_{23}-\widetilde\Theta_{23}) &=&\frac1{4ab}
(ab-a+b)(ab+a-b)\ . \nonumber
\end{eqnarray}

It is now clear that at the end we shall be able to get just one
relation for the three physical mixing angles. Indeed, Eqs. (\ref{e7}) 
and similar relations for the angles $\Theta_{12}$ and
$\widetilde\Theta_{12}$ (see below, Eq. (\ref{e15})) express all 
mixing angles (3 of them physical) through two unknown parameters 
$|A/C|$ and $|B/C|$.
However, it is necessary first to connect the physical angles entering
the CKM matrix with $\Theta_{23}$, $\widetilde\Theta_{23}$,
$\Theta_{12}$, $\widetilde\Theta_{12}$.

 One can easily get the relations between the "standard" angles 
$\vartheta_{12}$, $\vartheta_{23}$, $\vartheta_{13}$ in (\ref{CKM}) 
and
$\Theta_{23},\widetilde\Theta_{23},\Theta_{12},\widetilde\Theta_{12}$.
Using our definition of $K$, Eq. (\ref{b5}), and taking into account 
that $S_{12}=I$ (i.e. $\omega=0$) according to our solution of the
equation $d{\cal V}_{\rm eff}/d\omega=0$, one obtains:
\begin{eqnarray}\label{e8}
&&
s_{12}= \frac{\sin\Theta_{12}\cos
\widetilde\Theta_{12}\cos(\Theta_{23}-\widetilde\Theta_{23})
-\cos\Theta_{12}\sin\widetilde\Theta_{12} }{\sqrt{1-\sin^2\Theta_{12}
\sin^2(\Theta_{23}-\widetilde\Theta_{23})}} \approx 
\sin(\Theta_{12}-\widetilde\Theta_{12})     \nonumber \\
&& 
s_{23}=\frac{\cos\Theta_{12}
\sin(\Theta_{23}-\widetilde\Theta_{23})}{\sqrt{1-\sin^2\Theta_{12}
\sin^2(\Theta_{23}-\widetilde\Theta_{23})}} \approx 
\cos\Theta_{12}\sin(\Theta_{23}-\widetilde\Theta_{23}) \nonumber \\
&&
s_{13}=\sin\Theta_{12}\sin(\Theta_{23} - \widetilde\Theta_{23}) 
\end{eqnarray}
where we have approximated 
$\cos(\Theta_{23}-\widetilde\Theta_{23})\approx1$ and neglected 
$\sin^2\Theta_{12}\sin^2(\Theta_{23} -\widetilde\Theta_{23})\approx0$. 
% we get the simplified relations:
%\begin{eqnarray}
%s_{12}&=&\sin(\Theta_{12}-\widetilde\Theta_{12}\ , \nonumber \\
%s_{23}&=&\cos\Theta_{12}\sin(\Theta_{23}-\widetilde\Theta_{23})\ ,\\
%s_{13}&=&\sin\Theta_{12}\sin(\Theta_{23}-\widetilde\Theta_{23})\ .
%\nonumber
%\end{eqnarray}
%In Eq.(6.6) 
We have kept $\cos\Theta_{12}$ since the angle
$\Theta_{12}$ is slightly bigger than
$\Theta_{23}-\widetilde\Theta_{23}$:
$$ \Theta_{12}\simeq\tan\Theta_{12}=\frac{s_{13}}{s_{23}}\sim0.1,
\qquad \Theta_{23}-\widetilde\Theta_{23}=s_{23}\sim0.04\ . $$

We can now resolve the last of the equations (\ref{e7}) using the
smallness of $\sin^2(\Theta_{23}-\widetilde\Theta_{23})
=s^2_{23}/c^2_{12}$. We assume that this smallness is ensured by the
relation
%\begin{equation}
$ab-a+b\ \approx\ 0$,
%\end{equation}
which also leads to the smallness of $\Theta_{23}$ and
$\widetilde\Theta_{23}$ separately. In the linear approximation in
$s^2_{23}$ one has:
\begin{equation}\label{e9}
ab-a+b\ =\ 2\left(\frac{s^2_{23}}{c^2_{12}}\right) ~~~ \to ~~~ 
%\end{equation}
%or
%\begin{equation}
b\ =\ \frac1{a+1}\left[a+2\left(\frac{s^2_{23}}{c^2_{12}}
\right)\right]\simeq\frac a{a+1}\ .
\end{equation}
We shall see below that $a\sim 50$ while
$2s^2_{23}/c^2_{12}\sim3\cdot10^{-3}$. Thus, the second term in 
the square braclets in (\ref{e9}) is
completely negligible as compared to $a$.

Eq. (\ref{e9}) is the only physical consequence of the minimum of
the potential (\ref{e6}). According to Eq. (\ref{e6}) the "trivial" 
(vanishing) mixing angles, $\Theta_{23}=\widetilde\Theta_{23}=0$,
correspond to
\begin{equation}\label{e10}
\left[\frac{\cal V_{\rm eff}}{2|C|m^2_tm^2_b}\right]_{\rm trivial}=\ 
-a+b-1\ .
\end{equation}
whereas for the solutions (\ref{e7}) one gets:
\begin{equation}\label{e11}
\left[\frac{\cal V_{\rm eff}}{2|C|m^2_tm^2_b}\right]_{\rm non-trivial} =\
-a+b+1-\frac1{2ab}(ab-a+b)^2\ .
\end{equation}
Thus the non-trivial minimum is deeper than the trivial one. 
%value of $V_{eff}$, Eq.(6.10).

In order to find the angles $\Theta_{12}$ and $\widetilde\Theta_{12}$, 
we consider the next terms in (\ref{e4}):
\begin{eqnarray}\label{e12}
{\cal V}_{\rm eff}(\Theta_{12},\widetilde\Theta_{12}) = 
\left[Am^2_c\sin^2\Theta_{23} + Cm^2_cm^2_b\sin^2(\Theta_{23} - 
\widetilde\Theta_{23})\right] \cos^2\Theta_{12}  \nonumber \\
+ \left[Bm^2_s\sin^2\widetilde\Theta_{23}+Cm^2_tm^2_s\sin^2(\Theta_{23} -
\widetilde\Theta_{23})\right] \cos^2\widetilde\Theta_{12} 
\nonumber \\
+ Cm^2_cm^2_s \cos^2(\Theta_{12}-\widetilde\Theta_{12}) \ ,
\end{eqnarray}
where we have approximated $\cos(\Theta_{23}-\widetilde\Theta_{23})
=1$ in the last term of (\ref{e4}).
Substituting here $\Theta_{23}$ and $\widetilde\Theta_{23}$ from 
(\ref{e7}), one obtains
$$ \frac{{\cal V}_{\rm eff}(\Theta_{12},\widetilde\Theta_{12})} 
{2|C|m^2_cm^2_s}\ =\ -a'\cos2\Theta_{12}+b'\cos2
\widetilde\Theta_{12}-\cos2(\Theta_{12}-\widetilde\Theta_{12})\ , $$
with 
\begin{equation}\label{e13}
a'=\ \frac{m^2_b}{m^2_s}\ \frac1{2b}(ab-a+b) = \frac{\lambda_b}b\ ,
\qquad 
b'=\ \frac{m^2_t}{m^2_s}\ \frac1{2a}(ab-a+b) = \frac{\lambda_t}a\ ,
\end{equation}
where from Eq. (\ref{e9}) we have: 
\begin{equation}\label{e14}
%a'=\ \frac{\lambda_b}b\ , \qquad b'=\ \frac{\lambda_t}a\ ,
 \lambda_b=\ \left(\frac{m^2_b}{m^2_s}\right)\left(\frac{
s^2_{23}}{c^2_{12}}\right)\ , \qquad \lambda_t=\
\left(\frac{m^2_t}{m^2_s}\right)\left(\frac{s^2_{23}}{c^2_{12}}
\right)\ . 
\end{equation}

These equations show that, as anticipated, the two unknown
parameters $a'$ and $b'$ are expressed through one unknown number
$a$, or $b$, ($a$ and $b$ are not independent since are connected by 
Eq. (\ref{e9})) and physical mixing angles.

The angles $\Theta_{12}$ and $\widetilde\Theta_{12}$ can be now found
through $a'$ and $b'$ exactly in the same form as $\Theta_{23},
\widetilde\Theta_{23}$ through $a$ and $b$ (Eq. (\ref{e7})):
\begin{eqnarray}\label{e15}
\sin^2\Theta_{12}&=&\frac1{4a'^2b'}\ (a'b'-a'+b')(a'+b'-a'b')\ ,
\nonumber \\
\sin^2\widetilde\Theta_{12}&=&\frac1{4a'b'^2}\ (a'b'-a'+b')(a'
+b'+a'b')\ , \\
\sin^2(\Theta_{12}-\widetilde\Theta_{12})&=&\frac1{4a'b'}\
(a'b'-a'+b')(a'b'+a'-b')\ . \nonumber
\end{eqnarray}
We again can use the smallness of $\Theta_{12},\widetilde\Theta_{12}$
to assume that $a'b'-a'+b'\ =\ 0$.
Substituting here $a'$ and $b'$ from (\ref{e13}) and using the relation
(\ref{e9}), we get the equation for $b$. For the quantity $1-b$ this
equation has the form:
\begin{equation}\label{e16}
\lambda_t(1-b)^2-(\lambda_t\lambda_b+\lambda_t+\lambda_b)(1-b)
+\lambda_b=\ 0\ .
\end{equation}
It is easy to see that $b$ is close to unity. Indeed,
$\lambda_t\gg\lambda_b$, namely $\lambda_t\sim 35$ and
$\lambda_b\sim 2.5$. Choosing the proper sign for the root of
(\ref{e16}) we get
\begin{equation}\label{e17}
1-b =\frac1{2\lambda_t}\left[\lambda_t\lambda_b+\lambda_t +\lambda_b
-\sqrt{(\lambda_t\lambda_b+\lambda_t+\lambda_b)^2
-4\lambda_t\lambda_b}\right] 
%\nonumber\\
\simeq \frac{\lambda_b}{\lambda_t(1+\lambda_b)}\ .
\end{equation}
Thus $b$ differs from 1 only by a small correction:
\begin{equation}\label{e18}
b=1-\Delta, \qquad \Delta=\frac{m^2_bm^2_c}{m^2_sm^2_t}\cdot
\left(1+\frac{m^2_b}{m^2_s}\frac{s^2_{23}}{c^2_{12}}\right)^{-1}
\simeq\ 0.02\ .  
\end{equation}
The parameter $a$ is indeed large, $a=b/1-b\simeq50$, while the
parameters $a'$ and $b'$ are
\begin{equation}\label{e19}
a'=\frac{m^2_b}{m^2_s}\ s^2_{23}(1+\Delta)\simeq2.6\, \qquad
b'=\frac{m^2_t}{m^2_c}\ s^2_{23}\Delta\simeq0.7\ .
\end{equation}
It is straightforward now to derive the relation between the physical
mixing angles. From (\ref{e8}) and (\ref{e15}) we get
\begin{equation}\label{e20}
\frac{\sin^2(\Theta_{12}-\widetilde\Theta_{12})}{\sin^2\Theta_{12}}
=\frac{s^2_{12}(s^2_{23}+s^2_{13})}{s^2_{13}}=a'
\frac{a'+a'b'-b'}{a'-a'b'+b'}\ .
\end{equation}
This is already the sought relation because $a'$ and $b'$ are
already expressed through the mixing angles, Eqs. (\ref{e18}) 
and (\ref{e19}).
To present this relation in a more transparent way we use the
constraint $a'b'-a'+b'\approx 0$. Slightly more accurately it reads:
\begin{equation}\label{e21}
b'=\frac1{a'+1}(a'+2s^2_{12})\ .
\end{equation}
(Note that the accuracy of $a'b'-a'+b'= 0$ was quite adequate for 
the previous estimates. Its change to (\ref{e21}) leads only to a small 
change of $\Delta=1-b$ in Eq. (\ref{e17}): 
$\lambda_b\to\lambda_b+2s^2_{12}$).
Substituting (\ref{e21}) into (\ref{e22}) we obtain:
\begin{equation}\label{e22} 
\frac{s^2_{12}(s^2_{23}+s^2_{13})}{s^2_{13}}\ =\ a'^2\left[1+s^2_{12}
\left(1-\frac1{a'^2}\right)\right]\ .
\end{equation}
Or, with a linear accuracy in $s^2_{12}$, $(s_{13}/s_{23})^2$ and
$\Delta$:
\begin{equation}\label{e23}
\frac{s_{13}s_{23}}{s_{12}}(1+\Delta)\left[1+\frac12s^2_{12}\left(1-
\frac1{a'^2}\right)\right]\left(1-\frac12\frac{s^2_{13}}{s^2_{23}}
\right)=\frac{m^2_s}{m^2_b}\ .
\end{equation}
Here $\Delta$ and $a'$ are given by (\ref{e18}) and (\ref{e19}). 
Thus, neglecting the corrections which are of about $(3-4)\%$, the
final result comes out:\footnote{We have learned recently 
that the same relation was obtained in ref. \cite{Branco} 
in the completely different approach, namely by considering 
the mass matrix ansatzes with universal strength of the Yukawa 
couplings. We thank G. Branco for bringing his paper to our 
attention.} 
\begin{equation}\label{kiko}
\frac{s_{13}s_{23}}{s_{12}}\ =\ \frac{m^2_s}{m^2_b}\ .
\end{equation}
Neither the left-hand nor the right-hand side of this equation is
well known. However (\ref{kiko}) is satisfied within the present
experimental accuracy. Indeed, according to \cite{PDG}, we have 
$s_{12}=0.22$, $s_{23}=0.040\pm 0.005$ and 
$s_{13}/s_{23}=0.08\pm 0.02$. Substituting these in the 
left side of Eq. (\ref{kiko}), we obtain that 
$m_s/m_b=(2.4 \pm 0.6)\cdot 10^{-2}$, 
in agreement with the present understanding of the quark mass 
spectrum. 

% can lie within the interval $3\cdot10^{-4}$ to
%$10^{-3}$. For the masses $m_s=100-150$~MeV and $m_b=4.0-4.5$~GeV the
%ratio $(m_s/m_b)^2$ varies from $5\cdot10^{-4}$ to $1.4\cdot10^{-3}$.

%\begin{equation}
%\left|\frac{K_{ub}}{K_{cb}}\right|= 
%\frac{s_{13}}{s_{23}} = 
%\left(\frac{m_s/m_b}{s_{23}}\right)^2 s_{12} \ .
%\end{equation}

\section{Pseudo-familon masses }

The equation (\ref{kiko}) is the only relation between physical 
quantities following from the symmetry structure of the effective 
potential. On the other hand, the value of the effective potential, or 
more precisely, the value of its second derivative, determines the masses
of the "pseudo-\-fa\-mi\-lons" --- the pseudo-\-Gold\-stone bosons
corresponding to mixing angles. Though no reliable estimate of these
masses seems possible we shall try to suggest a guess of what they
could have been.

The typical pseudo-familon mass can be presented as
\begin{equation}\label{f1}
m^2_{PF}=\frac{d^2{\cal V}_{\rm eff}}{d\Phi^2_F}=
\frac1{F^2}\frac{d^2{\cal V}_{\rm eff}}{d\Theta^2}\sim
\frac1{F^2}{\cal V}_{\rm eff}.
\end{equation}
Here $\Phi_F$ is the pseudo-familon field, $\Theta$ and $F$ are the
corresponding angle and the scale, $\Phi_F=F\Theta$. The last
equality in (\ref{f1}) expresses the obvious fact that the 
differentiation
in angles do not change dramatically the value of the effective
potential.

For the estimate of the value of ${\cal V}_{\rm eff}$ we can use one 
of the diagrams of Figs. 5,6. We prefer to consider a three-loop 
diagram shown in Fig. 6 since it does not contain an unknown parameter
$\langle\Sigma\rangle$.   Approximately
\begin{equation}\label{f2}
{\cal V}_{\rm eff}\ \sim\ 
\frac{m_S^2}{(8\pi^2)^3 M^2}\ {\rm Tr} (\xi^+\xi\chi^+\chi) =
\frac{m_S^2M^2}{(8\pi^2)^3}\ {\rm Tr} (G_u^+G_uG_d^+G_d)\ ,
\end{equation} 
where $m_S\sim v$ is the supersymmetry breaking scale, 
%$\langle H^0_1\rangle\sim\langle H^0_2\rangle\sim v=246$~GeV
and $M$ is a mass of heavy fermions which in fact is a cutoff scale 
for quadratic divergency of this diagram. 
%$G_u=\chi/M$ and 
$G_u$ and $G_d$ are the MSSM Yukawa coupling matrices which 
are related to v.e.v.'s of $\chi$ and $\xi$ by Eqs. (\ref{c10}). 
We also introduce a factor $1/8\pi^2$ for each Feynman loop.

Let us consider for example pseudo-\-fa\-mi\-lons related to 
% $\Theta_{23},\widetilde\Theta_{23}$
23 (or 13) mixing angles. Then relevant terms in (\ref{f2}) 
are those involve the Yukawa constants of the third family: 
$G_t=\chi_3/M$ and $G_b=\xi_3/M$, and the relevant scale $F$ 
is essentially the smallest scale amongst $\chi_3$ and $\xi_3$, 
i.e. presumably $\xi_3$ if $G_b\ll G_t$ (The larger scale 
$\chi_3$ the corresponds to the "true" familon if the $SU(3)_H$ 
symmetry is global; it is absorbed by the horizontal gauge 
boson if $SU(3)_H$ is local). Then from Eqs. (\ref{f1}) and 
(\ref{f2}) we obtain:
%
%The v.e.v.'s of $\langle\chi\rangle$ and $\langle\xi\rangle$ are
%related to the masses of up- and down-\-quarks respectively
%(Eq. (\ref{c10})). Eq. (\ref{f2}) can be therefore rewritten in the form
%\begin{equation}\label{f3}
%{\cal V}_{\rm eff}\ \sim\ \frac{M^2_uM^2_d M^2}{(8\pi^2)^3 v^2}\ ,
%\end{equation}
%and, consequently:
%
\begin{equation}\label{f4}
m_{PF}(23)\sim\frac{G_tG_bm_SM}{(8\pi^2)^{3/2}\xi_3}\simeq  
\frac{G_tm_S}{16\sqrt{2}\pi^3} 
\end{equation}
which say for $m_S=700$ GeV is of the order of GeV. 
As for the familons related to the 12 mixing angles, one can obtain 
the similar estimate by taking the charm and strange quark 
constants $G_c$ and $G_s$ instead of $G_t,G_b$, and scale 
$\xi_2 \sim (G_s/G_b)\xi_3$. Therefore, these are lighter, 
with the mass of several MeV: $m_{PF}(12)\sim (m_c/m_t) m_{PF}(23)$.

Such massive pseudo-familons can decay into light quarks. 
For example, for 12 pseudo-familons with mass say 10 MeV 
we estimate the decay width 
into light quarks as 
\begin{equation}\label{decay}
\Gamma \sim \frac{1}{8\pi^2}\, 
\left(\frac{m_{u,d}}{F}\right)^2 \, m_{PF}(12)\ , ~~~~~  
\tau \sim \left(\frac{F}{10^{11}~{\rm GeV}}\right)^2 
\cdot 10^4 ~ {\rm s} 
\end{equation} 
since the ratio $m_f/F$, where $m_f$ is a fermion mass,
determines the strength of the familon coupling to fermion. 
Rather arbitrarily we have chosen the value $F=\xi_2\simeq 10^{11}$ GeV,
keeping in mind the typical value discussed usually for the breaking
of the Peccei-\-Quinn symmetry. Note also that $F$ cannot be much
smaller than $10^{10}$ GeV, due to the experimental bounds on the 
FCNC with the emission 
of the familon, like $K^+\to \pi^+ + {\rm familon}$  \cite{fam}.
It is easy to see that the lifetime of 23 pseudo-familons 
approximately scales with respect to that of 12 ones 
as inverse ratio of their masses, i.e. is smaller 
by about 2 orders of magnitude.

If pseudo-familons were in equilibrium in the early universe, 
then such long lifetimes ($\tau > 1$ s) 
%($\tau\sim 10^2-10^4$ s) 
can be somewhat problematic for nucleosynthesis. 
However, if the inflationary reheating temperature is 
considerably below the scale $F> 10^{10}$ GeV, which indeed 
seems to be the case e.g. due to constraint from the gravitino 
production, then familons would not be produced after the 
inflation.

\section{Discussion }

Summarizing the content of this work we would like to separate the
general idea which has been put forward from its concrete
implementation. The idea is that the weak mixing angles might be
actually the dynamical degrees of freedom --- the
pseudo-\-Gold\-stone bosons similar to the axion. The vacuum
expectation values of these fields fix the
Cabibbo--\-Kobayashi--\-Maskawa matrix. We believe that this general
assumption may survive even if the concrete scenario turns out to be
quite different from the one suggested in this paper. In that respect
what has been done may be considered as an existence proof.  
A clear lack of this model is that it cannot naturally explain 
the smallness of the angles $s_{12}$ and $s_{23}$ in the CKM matrix 
(e.g. in terms of the mass ratios), but rather implies certain 
fine tunings in adjusting their values to the experiment.  
At the same time we cannot help feeling a pleasant surprise that a 
model which we have chosen has led us to the relation (\ref{kiko}) 
which is in a reasonably good agreement with the experiment.

One interesting feature of our model is that $CP$-violating phase 
is vanishing in the CKM matrix. In other words, weak interactions 
cannot be responsible for the $CP$ violation in our model. 
However, $CP$ violation in the $K^0-\bar K^0$ system 
could emerge from the supersymmetric contributions to both 
$\epsilon_K$ and $\epsilon'_K$ parameters \cite{GG}, 
due to the flavour non-diagonal quark-squark-gluino 
couplings.\footnote{In this case typically one would obtain 
very small and maybe even negative $\epsilon'_K$, so that 
$CP$ violation in the $K^0-\bar K^0$ system could mimic the 
superweak mechanism.} 
Interestingly, the horizontal symmetry itself controlls that there 
can be no big flavour changing fermion-sfermion couplings to neutral 
gauginos.  In particular, if the horizontal $SU(3)_H$ symmetry is 
global, then the considered model satisfies the criteria given in 
ref. \cite{FC} and thus no flavour-changing effects would emerge 
at all beyond the usual MSSM ones. 
(The latter could not induce 
CP-violation in $K^0-\bar K^0$ system once the CKM matrix is real.) 
Nevertheless, if the $SU(3)_H$ symmetry is local, the flavour changing 
and CP-violating effects could be induced by the D-terms 
contributions \cite{HKR}. 
Alternatively, one could introduce some additional fermion states 
heavier than $U$ and $D$. Then some flavour changing effects could 
emerge at their decoupling \cite{DP}. 

Coming back once again to our initial point, if only the structure 
(\ref{a5}) emerges in the effective potential, then the CKM mixing 
angles are trivial. In order to deviate them from zeroes, 
some other terms should be introduced. In particular, we have 
included additional terms in the form of Eq. (\ref{b1}). 
However, the additional terms in principle 
could have completely different structures. 
For example, in the context of the left-right symmetric models 
one can imagine the situation when the structure (\ref{a6}) emerges 
in effective potential as well, while $K_L$ and $K_R$ are related 
through certain symmetry relations (in other words, left- and 
right rotation angles are not independent but do not coincide). 
In this case one could obtain a natural solution. 
Another possibility can be related to the grand unification theories, 
which introduce leptons into the consideration 
and thus could create alternative structures. 
The renormalization group effects could be also important for 
obtaining the non-trivial mixing angles. 

\vspace{1cm} 
{\bf Acknowledgments} 
\vspace{2mm} 

We would like to thank  A. Johansen, S.S. Gershtein, 
O. Kancheli, A. Masiero, R. Rattazzi, L. Randall and Q. Shafi 
for interesting discussions, and G. Dvali for informing us about 
the model based on similar ideas. 
We also thank participants of the PNPI 
theoretical seminar for useful comments. 
A.A. thanks G. Fiorentini, L. Piemontese and INFN section of Ferrara 
for their hospitality in Ferrara where this work has been done.

\newpage 
{\bf \large Figure Captions}
\vspace{1cm}

Fig. 1. The diagram due to the $W$ boson exchange which generates 
the term (\ref{a7}). \\ 

Fig. 2. Diagrams contributing the vacuum energy due to the Yukawa 
interactions, and leading to a structure (\ref{a5}).  \\

Fig. 3. An example of diagram with the charged Higgs exchange. \\  

Fig. 4. Diagrams generating the fermion masses via exchange 
of the heavy fermions $U$ and $D$. \\

Fig. 5. Supergraphs generating the first two terms in (\ref{c11}) 
after supersymmetry breaking. 
Insertions of spurions $z,\bar{z}$ are not shown. \\

Fig. 6. Supergraph generating third term in (\ref{c11}). 
Under the non-renormalizable vertices the tree-level graphs 
of Fig. 4 are understood.  \\

Fig. 7. Stable configurations of vectors with interactions given 
by Lagrangian ({\ref{d5}): (A) the case of three repulsions,    
$a,b,c>0$; (B) the case of two attractions and one repulsion, 
$a,b<0$, $c>0$. \\

Fig. 8. Contours restricting the parameter regions with the 
non-trivial minimum. Regions I and III correspond to the case $c=1$, 
and regions II and IV -- to the case $c=-1$. 

 \end{document}